\documentclass[sigconf]{acmart}
\usepackage{amsmath}
\usepackage{subfigure}
\usepackage{algorithm}
\usepackage{algorithmic}
\usepackage[utf8]{inputenc}

\usepackage{balance}
\usepackage{colortbl}
\usepackage{tabstackengine}

\AtBeginDocument{%
  \providecommand\BibTeX{{%
    \normalfont B\kern-0.5em{\scshape i\kern-0.25em b}\kern-0.8em\TeX}}}

\copyrightyear{2022}
\acmYear{2022}
\setcopyright{acmcopyright}
\acmConference[MM '22] {Proceedings of the 30th ACM International Conference on Multimedia }{October 10--14, 2022}{Lisboa, Portugal.}
\acmBooktitle{Proceedings of the 30th ACM International Conference on Multimedia (MM '22), October 10--14, 2022, Lisboa, Portugal}
\acmPrice{15.00}
\acmISBN{978-1-4503-9203-7/22/10}
\acmDOI{10.1145/3503161.3548130}
\begin{document}

\title{The Beauty of Repetition in Machine Composition Scenarios}

\author{Zhejing Hu}
\affiliation{\institution{The Hong Kong Polytechnic University}\country{Hong Kong SAR, China}}
\email{cszhu@comp.polyu.edu.hk}
\author{Xiao Ma}
\affiliation{\institution{The Hong Kong Polytechnic University}\country{Hong Kong SAR, China}}
\email{xiao1ma@comp.polyu.edu.hk}
\author{Yan Liu}
\authornote{Corresponding author}
\affiliation{\institution{The Hong Kong Polytechnic University}\country{Hong Kong SAR, China}}
\email{csyliu@comp.polyu.edu.hk}
\author{Gong Chen}
\affiliation{\institution{The Hong Kong Polytechnic University}\country{Hong Kong SAR, China}}
\email{csgchen@comp.polyu.edu.hk}
\author{Yongxu Liu}
\affiliation{\institution{The Hong Kong Polytechnic University}\country{Hong Kong SAR, China}}
\email{csyxl@comp.polyu.edu.hk}
\renewcommand{\shortauthors}{Zhejing Hu et al.}

\begin{abstract}
Repetition, a basic form of artistic creation, appears in most musical works and delivers enthralling aesthetic experiences. However, repetition remains underexplored in terms of automatic music composition. As an initial effort in repetition modelling, this paper focuses on generating motif-level repetitions via domain knowledge–based and example-based learning techniques. A novel repetition transformer (R-Transformer) that combines a Transformer encoder and a repetition-aware learner is trained on a new repetition dataset with 584,329 samples from different categories of motif repetition. The Transformer encoder learns the representation among music notes from the repetition dataset; the novel repetition-aware learner exploits repetitions' unique characteristics based on music theory. Experiments show that, with any given motif, R-Transformer can generate a large number of variable and beautiful repetitions. With ingenious fusion of these high-quality pieces, the musicality and appeal of machine-composed music have been greatly improved.

\end{abstract}



\maketitle

\section{Introduction}
Repetition, an act or an instance of repeating or being repeated, is ubiquitous in everyday life. It manifests in DNA sequencing; the Earth’s rotation; and activities such as swinging, exercising, and writing poems. It further influences people’s daily lives through the changing of seasons, tidal variation, and so forth. 

\begin{figure*}[htb]
	\centering
	\includegraphics[width=0.9\textwidth,trim=10 100 50 0,clip]{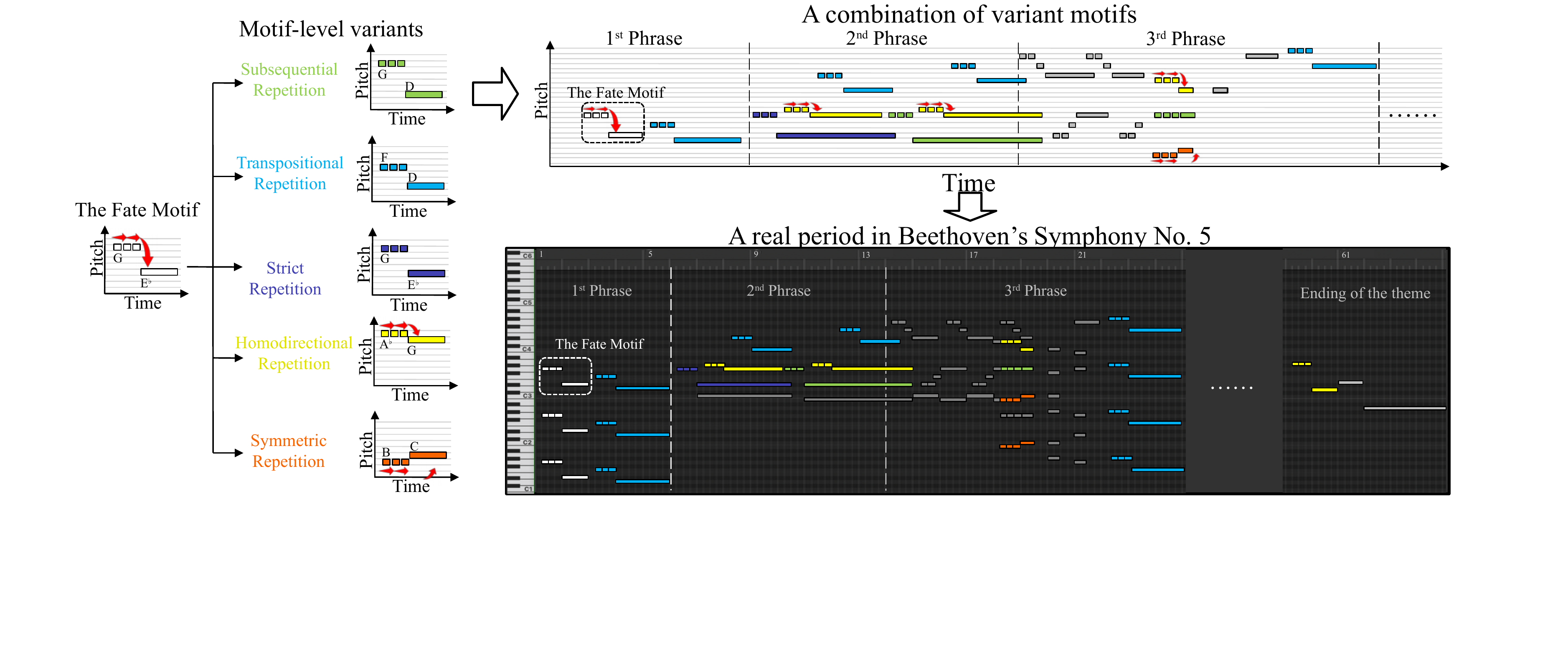}
  \caption{Beethoven's Fifth Symphony (simplified). The first four-note is the most recognizable ``fate motif''.}	
	\label{intro}
\end{figure*}

In music world, repetition and its variants generate  musical phantasmagoria with thrilling effects. Let's begin our exploration of this simple composition skill from the appreciation of a well-known classical work, Beethoven's Fifth Symphony (Figure \ref{intro}). An eighth note at the pitch of ``G'' with its two repetitions rings out, breaking the stillness, delivering a strong and firm signal. Then, a half note at the pitch of ``E-flat'' ends the motif succinctly that leaves the listener with a sense of breathlessness, dread and anticipation for the music to come. The ``G''s and the ``E-flat'' make up a descending Major 3rd interval and this is the famous fate motif, which delivers a feeling of ``fate knocking at the door''. Then, a transpositional repetition (TrR) of the fate motif that is the same motif repeats in a lower tone makes the atmosphere more solemn and dignified. In the second phrase, a strict repetition (StR) of the fate motif appears again with the same pitch value but longer ``E-flat'', leading to a smooth expansion of spectacular views. A homodirectional repetition (HoR) of triple ``A-flat'' and ``G'' triggers a precarious feeling. A subsequential repetition (SuR) ``G-G-G-D'' occurs with a long ``D'' note, forms a similar but lower ``platform'' like the strict repetition of the fate motif, embodying a groaning of the spirit. In addition, a transpositional repetition reaches a higher pitch, making the atmosphere more intense and enormous. In the third phrase, after a few alternating downward and upward notes, the appearance of a symmetric repetition (SyR) offers an upward movement, making a contrast that is distinct compared with other motifs. Three variants, a homodirectional repetition, a subsequential repetition, and a symmetric repetition, ring out together, forming a sense of contraction, bringing the piece to a tranquil state, symbolizing a truce, and heralding the arrival of a greater storm.

Music from every genre, culture, and period employs repetition for effect, yet these patterns have rarely been studied in machine composition \cite{2007Repetition}. One of music’s most repetition-obsessed composers, Steve Reich, exemplified this phenomenon in ``The Desert Music'' \cite{1989Steve}. Repetition in music does not simply entail identical elements but rather echoing elements in a new way \cite{fink2005repeating}. In other words, patterns of repetition are more than composition skills—they invite listeners to participate.

Repetition can be categorized along multiple dimensions since it is diverse, complicated, and includes different composition forms. First, different criteria can be used to classify repetition (e.g., pitch values, rhythm, and harmony) \cite{dai2020automatic,zou2021melons}. To avoid overlap due to different criteria, we focus on the in-depth deconstruction of musical repetition in terms of pitch value (Table \ref{tab:repetition}). Second, repetition can be divided by level—the note level, motif level, phrase level, and period level. As the most basic analyzable and meaningful element of a musical composition, motifs create dynamic coherence through repetition, variation, and combination at different scales \cite{1990Music}. Therefore, motif-level repetitions are examined in this paper. 

The diversity and complication of repetition in music further results in two challenges. First, the scarcity of repetition-related data precludes investigation of different repetition types. No public dataset is available to specify repetition types; it is accordingly difficult to further analyze music repetition. Second, how to produce and combine explicit repetitions to make pleasant music is still challenging. Currently, some work uses rule-based and statistical methods to construct long-term repetitive structures \cite{elowsson2012algorithmic,dai2021personalized,collins2017computer,2011Music,2017MorpheuS}. Others employ memory modules in deep learning models, such as long short-term memory (LSTM) and Transformer, to generate music \cite{mangal2019lstm,huang2018music,huang2020pop,wu2020popmnet,hsiao2021compound,dai2021controllable}. However, these models cannot generate explicit repetition types since they lack the domain knowledge of music repetition types. It is hard to understand and follow the structure and direction of the generated music.
 
To surmount these obstacles, we first present a new dataset, Music Repetition Dataset (MRD), with repetition labels based on these definitions. We also develop a novel model called Repetition Transformer (R-Transformer) that can be controlled to generate designated music repetition given a music piece. This model combines music knowledge and a Transformer encoder. The domain knowledge-based mechanism allows the model to explicitly follow music theory while the transformer encoder mechanism enables the model to learn the representation of notes. In this model, a novel repetition-aware learner is designed to generate different repetitions based on their unique characteristics learned from a Transformer encoder. Based on multi-aspect attribute representation in music repetition, parameters in repetition-aware learner can be controlled to exploit unique characteristics of different repetitions. To better produce different repetition types, a reconstruction and classification cooperation method is proposed to train R-Transformer. In generation phase, a rule-based generation module is applied to generate the designated repetition given an repetition type, which helps users control the process of composing different repetition types. They can thus create music more precisely.

\begin{table*}[htb]
\centering
	\setlength\tabcolsep{1.5pt}
  \caption{Types of repeated patterns in terms of pitch.}

  \label{tab:repetition}

  \begin{tabular}{ccll}
    \toprule
   Scale&No. &Repetition Name&Definition\\
    \midrule
Note&1&Same note&Two notes are the same.  \\
&2&Same pitch class&Two notes are in the same pitch class.\\
\hline
Motif&3&Strict repetition ((StR)&
Two motifs are the same. \\
&4&
\Centerstack{Transpositional repetition (TrR)}\footnotemark[1]& Two motifs have the same relationships between pitches but with a different tone. \\
&5&
\Centerstack{Subsequential repetition (SuR)}&
Two motifs are not identical or transpositional, but share a common subsequence. \\
&6&
\Centerstack[c]{Homodirectional repetition (HoR)}&Two motifs are not SuR, but have homodirectional repetition development\footnotemark[2].\\
&7&
\Centerstack{Symmetric repetition (SyR)}&
Two motifs are not SuR, but have similar symmetric repetition development. \\
\hline

Phrase&8&Similar melody&Two phrases have similar melody. \\
Period&9&Structural repetition& Two periods have similar structure.\\
  \bottomrule
\end{tabular}
\end{table*}

The rest of this paper is organized as follows. In Section 2, we review related work on music repetition generation. In Section 3, we describe the proposed method in detail. Sections 4 and 5 discuss the implementation details and experimental results. We close with our conclusions and directions for future work.

\section{Related Work}

Different from rule-based \cite{dai2021personalized} or convolutional neural network (CNN)–based generation models \cite{dong2018musegan,yang2017midinet,chen2018musicality}, Transformer \cite{vaswani2017attention} has been widely adopted as the backbone generative model: the ``self-attention'' mechanism can provide a much longer memory so the model can learn the repetitive structure of music. Music Transformer \cite{huang2018music} was the first Transformer-based model in symbolic music generation, showing that a Transformer model can generate coherent minute-long polyphonic piano music with reasonable repetitions and variance. Other Transformer-based models have since been proposed to generate music \cite{2020Transformer,zhang2020learning,muhamed2021symbolic,ens2020mmm,ren2020popmag,huang2020pop,hsiao2021compound}. These models can learn musical features without hand-crafted rules and produce diverse music pieces. However, without studying music repetition and analyzing the music structure, the model tends to lose a specific sense of direction \cite{hernandez2021music} and the generated music tends to be random without rhythmic patterns \cite{wu2020jazz}. Overall, machine learning–based models continue to struggle to adhere to certain key musical ideas in composition and to generate repetitive rhythmic patterns.

Scholars have sought to analyze the hierarchical structure of music to generate repetitive music patterns \cite{collins2017computer,medeot2018structurenet,jhamtani2019modeling,wu2019hierarchical}. Recently, MusicFramworks \cite{dai2021controllable} was proposed to use a Transform and LSTM-based model to create a full-length melody guided by a long-term repetitive structure. PopMNet \cite{wu2020popmnet} involves a structured generation network and a melody generation network based on the structural representation and chord progression of input music. MELONS \cite{zou2021melons} entails a multi-step generation method with Transformer-based models and graph representation for music structure generation and structure-conditional melody generation. Theme Transformer \cite{shih2022theme} achieves theme-based conditioning by producing it multiple times in the generation result so that the output music follows the thematic material. These models were constructed based on the structure of music from multiple levels (e.g., the motif level and phrase level). Generating music based on the structure of the music can relieve the problem that generated music might lose a specific sense of direction. Yet none of these models studies music repetition modeling and they cannot generate explicit repetition types. Without modeling music repetition, users cannot control repetition and music trend. Generating repetitive patterns and understandable music rhythmically is still a challenge in these models. 

\footnotetext[1]{Both chromatic and diatonic transposition are considered. Chromatic is scalar transposition within the chromatic scale, implying that every pitch in a collection of notes is shifted by the same number of semitones. Diatonic transposition is scalar transposition within a diatonic scale (a standard scale under some tonality indicated by a certain standard key signature).}
\footnotetext[2]{Development: the direction of the pitch sequence from one note to the next.}

\begin{figure*}[htb]
	\centering
	{\centering\includegraphics[width=0.95\linewidth,trim=0 100 0 0,clip]{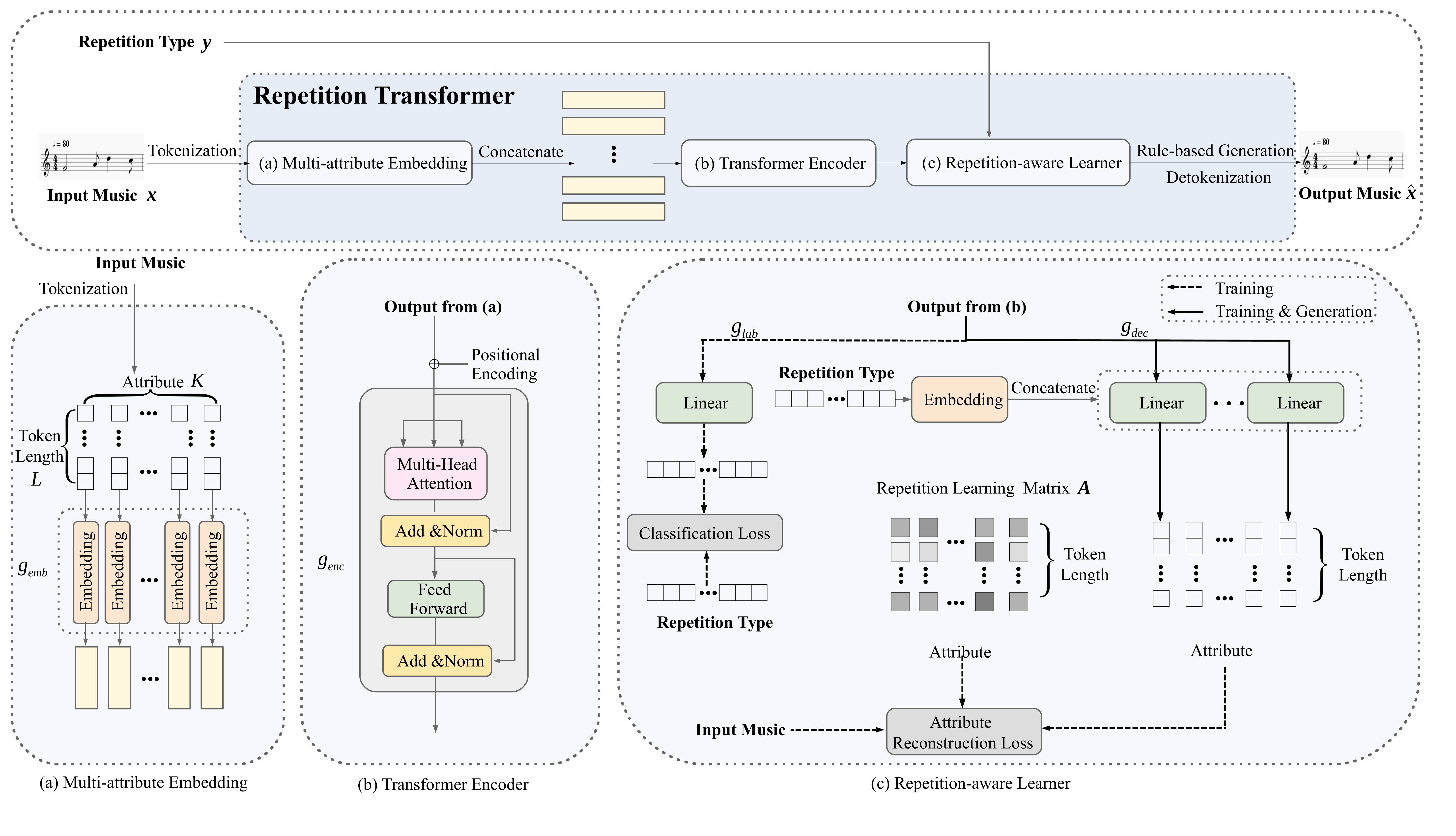}}
	\caption[.]{A general framework for R-Transformer. R-Transformer linearly embeds multi-aspect attributes and feeds them into a Transformer encoder. We define a repetition-aware learner that can account for the contributions of different attributes. We employ one classification and one reconstruction loss to train the model.	}	
	\label{framework_train}
	\vspace{-5pt}
\end{figure*}
\section{Method}

In this section, we introduce our R-Transformer architecture and elaborate on classification and reconstruction cooperation learning to train R-Transformer from scratch. Figure \ref{framework_train} provides an overview of the architecture. We feed the input music and its repetition type into the model, where the raw input is first tokenized and then embedded to multiple attribute-embedding vectors followed by a Transformer encoder and a repetition-aware learner. There are three major modules in R-Transformer: 1) a multi-attribute-embedding layer; 2) a single backbone Transformer encoder applied to all music attributes; and 3) a repetition-aware learner to learn each attribute separately. A reconstruction and classification cooperation training strategy is proposed in this paper to generate different repetition types.

Mathematically, given an input music repetition after tokenization $\mathbf{x} \in \mathbf{X}$ and a repetition type $\textbf{y} \in \mathbf{Y}$, where $\mathbf{X} \subseteq \mathbb{R}^{L \times K}$ is the input space and $\mathbf{Y} \subseteq \mathbb{R}^{M}$ is the label space. $L$ is the length of the input, $K$ is the number of music attributes such as note pitch, note duration, and etc., and $M$ is the number of repetition types. The goal is to generate a new piece of music $\hat{\textbf{x}} \in \hat{\mathbf{X}}$, where $\hat{\mathbf{X}} \subseteq \mathbb{R}^{L \times K} $.  
Let $f_{c}: \mathbf{X} \rightarrow \mathbf{Y}$ be the label learning pipeline and $f_{r}: \mathbf{X} \rightarrow \hat{\mathbf{X}}$ be the data reconstruction pipeline of R-Transformer. We define an embedding space $\mathbf{H} \subseteq \mathbb{R}^{L \times H_{1}}$ and a feature space $\mathbf{F} \subseteq \mathbb{R}^{L \times H_{2}}$, where $H_{1}$ and $H_{2}$ are embedding and feature dimensions. We also define four additional functions: 1) a multi-attribute-embedding mapping $g_{\text{emb}} : \mathbf{X} \rightarrow \mathbf{H}$; 2) an encoder/feature mapping $g_{\text{enc}}: \mathbf{H} \rightarrow \mathbf{F}$; 3) a decoder $g_{\text{dec}}: \mathbf{F} \rightarrow \hat{\mathbf{X}}$; 4) a feature-labeling function $g_{\text{lab}}: \mathbf{F} \rightarrow \mathbf{Y}$. 
The label learning pipeline $f_{c}$ and the reconstruction pipeline $f_{r}$ can be decomposed such that
\begin{equation}
    f_{c}(\textbf{x}) = (g_{\text{lab}} \circ g_{\text{enc}} \circ g_{\text{emb}})(\textbf{x}),
    \label{eqs1}
\end{equation}
\begin{equation}
    f_{r}(\textbf{x},\textbf{y}) = (g_{\text{dec}} \circ g_{\text{enc}} \circ g_{\text{emb}})(\textbf{x},\textbf{y}),
        \label{eqs2}
\end{equation}
where $\circ$ is the composition of two functions. In Equation \ref{eqs1}, the embedding $g_{\text{emb}}$ is a multi-attribute embedding module (Figure \ref{framework_train} (a)). The encoder $g_{\text{enc}}$ is a Transformer encoder (Figure \ref{framework_train} (b)). The labeling module $g_{\text{lab}}$ consists of a linear projection (Figure \ref{framework_train} (c)). In Equation \ref{eqs2}, the output of $g_{\text{enc}}$ is concatenated with a target repetition label $\textbf{y}$ and fed into $g_{\text{dec}}$, which consists of $K$ linear projection layers to predict $K$ attributes of the output music (Figure \ref{framework_train} (c)). 

\subsection{Multi-Attribute Embedding}
In this step, a multi-attribute embedding module is implemented to embed multi-aspect attributes of the music separately, which contain music information such as note pitch, note duration, and so on. The input music $\textbf{x}$ is embedded and the output of multi-attribute embedding $g_{\text{emb}}$ after concatenation is 
$
    [\text{Embed}_{1}(\textbf{x}_{:,1}),\cdots,\text{Embed}_{K}\\(\textbf{x}_{:,K}) ],$
where $\text{Embed}_{k}(\cdot)$ is the embedding layer of $k$-th attribute and $[\cdot,\cdot]$ is the action of concatenation.

\subsection{Transformer Encoder Architecture}
We adopt the most established Transformer encoder architecture \cite{vaswani2017attention}, which has been frequently applied in music generation as of late \cite{hsiao2021compound,zou2021melons}. We follow the standard Transformer encoder architecture \cite{vaswani2017attention} and illustrate it in Figure \ref{framework_train} (b). The sequence of input tokens to the Transformer encoder is $g_{\text{emb}}(\textbf{x}) + \mathbf{e}_{\text{pos}}$, where $\mathbf{e}_{\text{pos}} \in \mathbb{R}^{L\times H_{1}}$ is a positional embedding vector. The output in the Transformer encoder is used as the aggregated representation for the entire input sequence. This output will later be used for classification ($g_{\text{lab}}$) and reconstruction ($g_{\text{dec}}$). We employ a standard self-attention as the multi-head-attention module with GeLU \cite{vaswani2017attention} as the activation in the linear projection layer. We also use layer normalization \cite{ba2016layer} after the multi-head-attention and linear projection modules. 

\subsection{Repetition-aware Learner}
In this step, we propose a novel repetition-aware learner module to train the model and generate music. The output of $g_{\text{enc}}$ is fed into a feature-labeling function $g_{\text{lab}}$ to predict the repetition type $\textbf{y}$. $g_{\text{enc}}$ is also fed into a linear decoder $g_{\text{dec}}$ concatenated with  the target repetition label to reconstruct the input music $\textbf{x}$. 

Specifically, we first define a repetition learning matrix that enables us to directly control the training process following the definition of repetition types. Then, a reconstruction and classification cooperation learning strategy is proposed to train the model. 

\textbf{Repetition learning matrix}: It is known that different repetition types have different characteristics; and different motifs have different representations. To learn these differences, we study multi-aspect attributes by first defining a repetition learning matrix $\mathbf{A} \in \mathbb{R}^{L \times K}$. 

The element $a_{l,k}$ in repetition learning matrix $\mathbf{A}$ can be calculated as 
\begin{equation}
a_{l,k} = \gamma \cdot (1+\omega_{l,k}),
\label{a}
\end{equation}
where $\gamma$ is an attribute importance hyper-parameter that controls the importance of each attribute. $0\leq \omega \leq 1$ is determined by the appearance frequency of each element in the music and can be calculated as 
\begin{equation}
    \omega_{l,k} =\dfrac{Counter(x_{l,k}, \textbf{x}_{:,k})}{L},
    \label{omega}
\end{equation}
where $Counter(\cdot)$ is a function that counts the occurrence number of each element in each attribute.

The process of repetition-aware learning is depicted in Figure \ref{framework_train} (c). The main intuition behind the multi-aspect attribute learning is from two aspects. First, as described in Table \ref{tab:repetition}, different attributes represent differently in different types. Although pitch is the main attribute that determines repetition types, other attributes such as note duration, note position and note velocity also contribute differently. For example, strict repetition (StR) have same number of notes and pitch values are identical, while subsequential repetition (SuR) of two motifs might have different number of notes and different position, duration and velocity of notes. Therefore, in SuR, $\gamma$ in Equation \ref{a} will be given a higher attribute importance if $k$ is pitch, position, duration and velocity. Second, the representation of different motifs varies within the same repetition type. In other words, the distribution of each attribute is totally different among motifs. For example, in pitch dimension, ``C3-D3-E3-F3-G3'' and ``C3-D3-C3-E3-C3'' have different representations. If $k$ is pitch, then $a_{1,k}$, $a_{3,k}$ and $a_{5,k}$ will be larger based on Equations \ref{a} and \ref{omega} in the latter case, since ``C3'' appears more frequently than other pitch values.

\textbf{Learning algorithm:} Given labeled input music samples $\{(\textbf{x}_{i},\\ \textbf{y}_{i})\}_{i=1}^{N}$, where $\textbf{y}_{i} \in \{0,1\}^{M}$ is a one-hot vector with $M$ classes and $N$ is the number of music samples. Let $\theta_{c} = \{\theta_{\text{emb}},\theta_{\text{enc}},\theta_{\text{lab}}\}$ and $\theta_{r} = \{\theta_{\text{emb}},\theta_{\text{enc}},\theta_{\text{dec}}\}$ denote the parameters of the label learning pipeline $f_{c}$ and the data reconstruction pipeline $f_{r}$. $\theta_{\text{emb}}$ and $\theta_{\text{enc}}$
are shared parameters for the embedding $g_{\text{emb}}$ and feature mapping $g_{\text{enc}}$. Let $\ell_{c}$ and $\ell_{r}$ be the classification and reconstruction loss, respectively. Based on Equations \ref{eqs1} and \ref{eqs2}, we first define the empirical classification loss as
\begin{equation}
    \mathcal{L}_{c} ( \{\theta_{\text{emb}},\theta_{\text{enc}},\theta_{\text{lab}}\}):=
    \sum_{i=1}^{N} \ell_{c} (f_{c}(\textbf{x}_{i};\{\theta_{\text{emb}},\theta_{\text{enc}},\theta_{\text{lab}}\}),\textbf{y}_{i}).
\end{equation}
Typically, $\ell_{c}$ is of the form cross-entropy loss $\sum_{m=1}^{M}y_{k}log[f_{c}(\textbf{x})]_{m}$ (recall that $f_{c}(\textbf{x})$ is the softmax output). In addition, the attribute reconstruction loss is defined as
\begin{equation}
    \mathcal{L}_{r} ( \{\theta_{\text{emb}},\theta_{\text{enc}},\theta_{\text{dec}}\}):=
    \sum_{i=1}^{N} \ell_{r} (f_{r}(\textbf{x}_{i},\textbf{y}_{i};\{\theta_{\text{emb}},\theta_{\text{enc}},\theta_{\text{dec}}\}),\hat{\textbf{x}}_{i}).    
\end{equation}
$\ell_{r}$ is of the form squared loss combined with repetition learning matrix:
\begin{equation}
\ell_{r} = ||\textbf{A} \otimes  (\hat{\textbf{x}}-f_{r}(\textbf{x},\textbf{y}))||^{2}_{2},
\end{equation}
where $\otimes$ is the element-wise product of two matrices. 

The final objective of our proposed method to minimize the total loss $\mathcal{L}$ and is formulated as follows:
\begin{equation}
    \underset{\theta_{\text{emb}},\theta_{\text{enc}},\theta_{\text{dec}},\theta_{\text{lab}}}{\text{min}}\lambda  \mathcal{L}_{c} ( \{\theta_{\text{emb}},\theta_{\text{enc}},\theta_{\text{lab}}\}) +(1-\lambda)\mathcal{L}_{r} ( \{\theta_{\text{emb}},\theta_{\text{enc}},\theta_{\text{dec}}\}),
    \label{obj}
\end{equation}
where $0\leq \lambda \leq 1$ is a hyper-parameter controlling the trade-off between classification and reconstruction. The objective is a convex combination of supervised and unsupervised loss functions. 

The objective in Equation \ref{obj} can be achieved by minimizing $\mathcal{L}_{c}$ and $\mathcal{L}_{r}$ using stochastic gradient descent. The stopping criterion for the algorithm is determined by monitoring the average total loss during training – the process is stopped when the average total loss stabilizes. 

\begin{algorithm}[h]
	\caption{R-Transformer in the training \& generation phase.}
	\label{alg1}
	\begin{flushleft}
	\textbf{Training phase:}\\
	\textbf{Input}: Music data $\{(\textbf{x}_{i}, \textbf{y}_{i})\}_{i=1}^{N}$\\
	\textbf{Parameter}:  Learning rate $\alpha$, trade-off parameter $\lambda$ \\
	\textbf{Output}: Optimal parameters $\theta^{*} =\{ \theta^{*}_{\text{emb}},\theta^{*}_{\text{enc}},\theta^{*}_{\text{lab}}, \theta^{*}_{\text{dec}}\}$
    \end{flushleft} 
	\begin{algorithmic}[1] 
	    \STATE Initialize parameters  $\theta_{\text{emb}},\theta_{\text{enc}},\theta_{\text{lab}}, \theta_{\text{dec}}$
	    \WHILE {\text{not converge}}
	    \FOR {\textbf{each} batch of size $n$}
		\STATE  Do a forward pass $f_{c}(\textbf{x}) = (g_{\text{lab}} \circ g_{\text{enc}}\circ g_{\text{emb}})(\textbf{x});$
		\STATE Do a forward pass $f_{r}(\textbf{x},\textbf{y}) = (g_{\text{dec}} \circ g_{\text{enc}}\circ g_{\text{emb}})(\textbf{x},\textbf{y});$
        \STATE Calculate repetition learning matrix $\textbf{A}$ based on Equations \ref{a} and \ref{omega};
        \STATE Update $\theta:$
		    $\theta \leftarrow \theta -\alpha \bigtriangledown_{\theta} \mathcal{L}^{n}(\theta);$
        \ENDFOR
		\ENDWHILE
	\end{algorithmic}
	\begin{flushleft}
	\textbf{Generation phase:}\\
    \textbf{Input}: Music data $\textbf{x}$ and designated repetition type $\textbf{y}$\\
	\textbf{Parameter}: $\theta^{*}_{\text{emb}},\theta^{*}_{\text{enc}}, \theta^{*}_{\text{dec}}$\\
	\textbf{Output}: Output music $\hat{\textbf{x}}$
    \end{flushleft} 	
    \begin{algorithmic}[1]
        \IF {$\textbf{y}$ is StR}
			\IF {$k$ is pitch}
			    \STATE $\hat{x}_{l,k} = x_{l,k}$;
			\ENDIF
		\ELSIF {$\textbf{y}$ is TrR}
		    \IF {$k$ is pitch}
		        \STATE $\hat{x}_{l,k} = x_{l,k}+t$;
		    \ENDIF
		\ELSE
        \STATE  $\hat{\textbf{x}} = f_{r}(\textbf{x},\textbf{y}) = (g_{\text{dec}} \circ g_{\text{enc}}\circ g_{\text{emb}})(\textbf{x},\textbf{y})$;
	    \ENDIF
    \end{algorithmic}
\end{algorithm}
\textbf{Rule-based generation:} In the generation phase, only the reconstruction pipeline in the repetition-aware learner will be activated to generate the output, that is, $\hat{\textbf{x}} = f_{r}(\textbf{x},\textbf{y})$ where $\textbf{x}$ and $\textbf{y}$ are the input music and the designated repetition type. In addition, we also update the results by following the definition of different repetitions based on music theory. For instance, in a StR, the pitch value is assumed to be identical to the input motif, so we fix the pitch value. Specifically, if $k$ represents pitch, then $\hat{x}_{l,k} = x_{l,k}$. In a TrR, the difference of pitch value is assumed to be fixed, so we follow the rule to generate pitch value after the first pitch value is predicted. Specifically, if $k$ represents pitch, then $\hat{x}_{l,k} = x_{l,k}+t$, where $t =\{\cdots,-2,-1,1,2,\cdots\}$ is a transposition value. In SuR, HoR and SyR, the representation of different notes is learned by the model, so all music attributes are generated based on the example-based mechanism. Furthermore, user can generate multiple repetition types by giving the model multiple labels. For example, the model can generate four motifs sequentially if four labels are given. The training and generation phase of R-Transformer learning algorithm is summarized in Algorithm \ref{alg1}. In terms of StR and TrR, the proposed model utilizes rule-based generation to ensure the pitch attribute is identical to the input. In addition, example-based mechanism allows the proposed model to learn representations in terms of other attributes from music examples, which makes the output music diverse to hear. 

\section{Experimental Configurations}

We train our model using the Pop piano dataset from \cite{hsiao2021compound} because the structures in pop music are relatively simple. In addition, this dataset covers all repetition types shown in Table \ref{tab:repetition}. This dataset contains 1,748 pieces of pop piano from the Internet. All songs are in 4/4 time signature, and each song is converted into a symbolic sequence following transcription, synchronization, quantization, and analysis. 
\subsection{Music Repetition Dataset}

\begin{figure}[htb]
	\centering
	{\centering\includegraphics[width=0.95\linewidth,trim=19 81 18 159,clip]{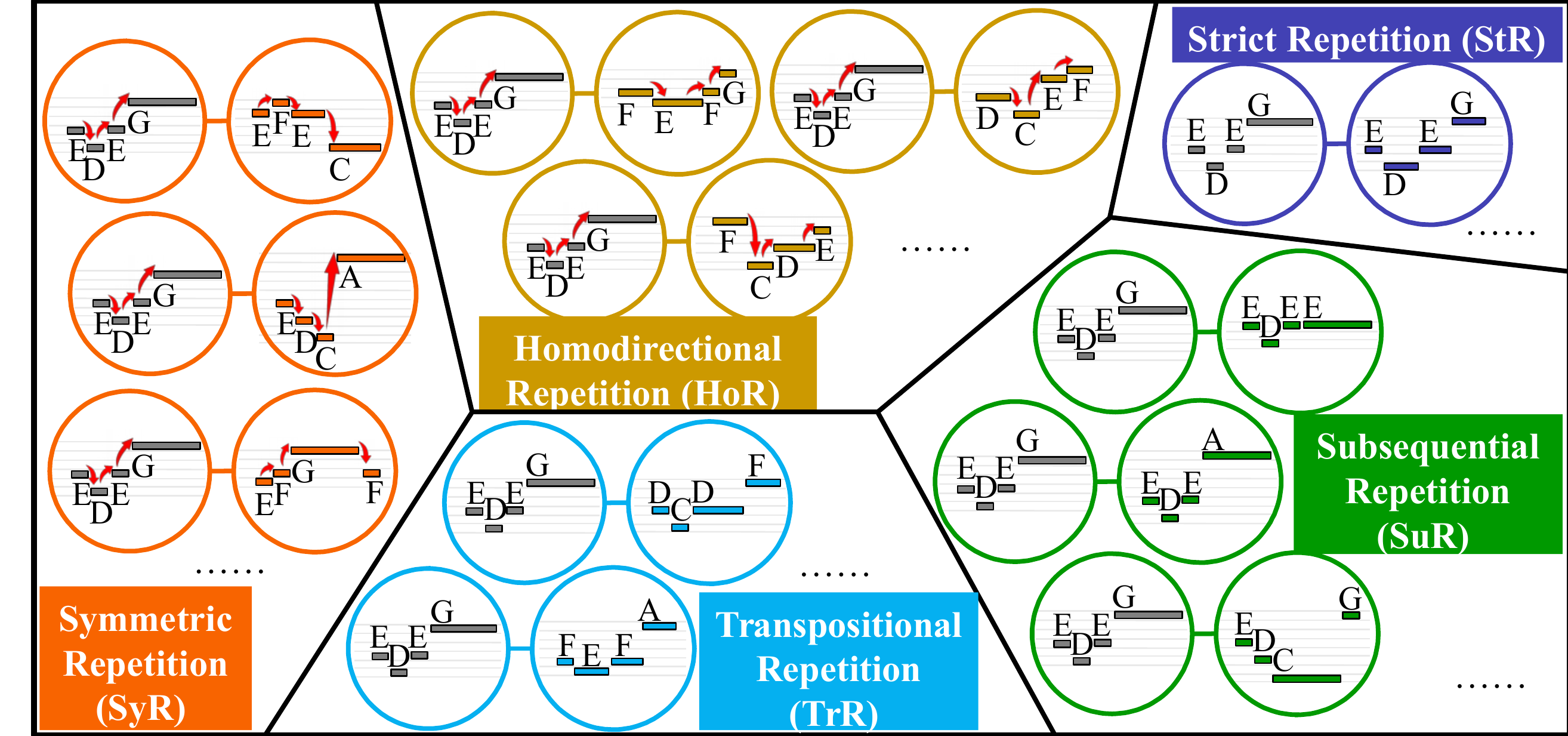}}
	\caption[.]{Examples of five repetition types in MRD.}	
	\label{motif}
	\vspace{-5pt}
\end{figure}

We further partition the music piece into motifs. After partition, we compare the pitch value of all motifs. Pitch and development are used since they are both commonly used in music creation, and can be easily recognized and understood by the audience \cite{2011On}. If two motifs satisfy the definition in Table \ref{tab:repetition}, then it is a repetition; otherwise, it is not a repetition. In StR, the pitch value of all notes should be identical. For other repetition types, the melody of the note sequence is compared if the music contains both melody and accompaniment. 
\begin{table}[h]
	\setlength\tabcolsep{2pt}
  \caption{Percentage of different repetition types in training and test dataset of MRD.}
  \label{tab:train}
  \begin{tabular}{ccccc}
    \toprule
    Dataset&Repetition&Numbers&Percentages&Avg Length (tokens)\\
    \midrule
Training&    StR &21807&3.88\% &64.93 \\
&TrR &27852& 4.95\% &62.94\\
&SuR &73774& 13.11\% &73.74\\
&HoR& 152060&27.02\% &59.68\\
&SyR& 287070&51.03\%& 64.62\\\hline

   Test &StR &1395&6.41\% &61.22 \\
    &TrR &1635& 7.51 \% &60.81\\
    &SuR &3331& 15.31\% &69.71\\
    &HoR& 5038&23.15\% &64.68\\
    &SyR& 10367&47.63\%& 68.97\\
    
  \bottomrule
\end{tabular}
\end{table}

In SyR, the development of two motifs exhibiting vertical, horizontal or rotational symmetry constitutes a symmetric repetition. Inversion is an extreme case in horizontal symmetry where ascending developments are made to descend by the same degree \cite{kempf1996symmetry}. For example, if the original motif is ``E4-D4-E4-G4'' (Down-Up-Up), then horizontal symmetry ``E4-F4-E4-C4'' (Up-Down-Down), vertical symmetry ``E4-F4-G4-F4'' (Up-Up-Down) and rotational symmetry ``E4-D4-C4-A4'' (Down-Down-Up) are symmetric repetition. In addition, in SuR, HoR, and SyR, we set the similarity of two pitch sequences or development sequences as 75 \%. We omit motifs that satisfy both HoR and SyR (e.g., ``C4-D4-E4'' and ``C4-E4-G4''), as this situation might lead to overlap between repetition types and is not prevalent in this dataset (less than 1\%). However, this type of repetition presents an intriguing topic for further analysis. Figure \ref{motif} illustrates some examples of five repetition types in MRD. First, they are exclusive based on the definition in Table \ref{tab:repetition}. Each motif is categorized into one repetition type. Second, the size of five repetition types is different. StR has the smallest size since the definition of StR is the most strict that all note pitches should be identical. SyR has the largest size since it measures the development of the music and contains three symmetric scenarios.

To retain most information from the motifs, we set the longest motif sequence in the dataset as the default length (i.e. 120 tokens). If the sequence is shorter than the max length, then we pad zeros to the sequence. In addition, we randomly hold out motifs from 100 songs for testing and use the remaining motifs to train R-Transformer. The basic statistics of five repetition types appear in Tables \ref{tab:train}.

\subsection{Implementation Details}
Following the backbone model of \cite{hsiao2021compound}, we also choose the linear Transformer \cite{katharopoulos2020transformers} for sequence modeling. The Transformer encoder module consists of 6 self-attention layers with 4 heads and 256 hidden states. The inner layer size of the feed-forward part is set to 2048. The symbolic music is converted to a sequence of compound words drawn from a pre-defined music attribute vocabulary \cite{hsiao2021compound}. Musical information is explained based on seven attributes, tempo, chord, position, type, pitch, duration, and velocity. The embedding sizes of these seven attributes and label are 128, 256, 64, 32, 512, 128, 128, and 32, respectively. The embedding sizes are chosen based on the vocabulary sizes of different token types, and embedded tokens describing the same object are concatenated together and linearly projected to the same size of the corresponding module’s hidden state. We use the Adam optimizer \cite{kingma2014adam} with a learning rate of $10^{-4}$ for the reconstruction and classification model. We apply dropout with a rate of 0.1 during training. In our model, repetition importance factors are designed as follows: in all types, $\gamma$ of pitch is 4 since pitch contributes more to the repetition type. In SuR, HoR and SyR, $\gamma$ of position, duration and velocity is 2 since these contributes also determines the representation of notes. For other attributes, $\gamma$ is set as 1. $\lambda$ in Equation \ref{obj} is 0.5.

\section{Experiment}

\subsection{Model Evaluation and Comparison}

In both objective and subjective experiments, five relative models are adopted for comparison: \textbf{CP-C} and \textbf{CP-NC} are two versions from the state-of-the-art music generation model \cite{hsiao2021compound}. \textbf{CP-C} generates the new music by taking one motif as the input condition. \textbf{CP-NC} generates the music without any conditions and the matching rate is calculated based on the first generated motif. \textbf{PopMNet}, \textbf{Theme} and \textbf{MELONS} generate music by studying the repetitions and variance in the music structure. \textbf{PopMNet} uses a combination of recurrent neural network (RNN) and CNN models. \textbf{Theme} uses Transformer-based models. \textbf{MELONS} uses Transformer- and graph-based models. 

In the objective evaluation, we design two additional baseline models \textbf{R-Transformer-V} and \textbf{R-Transformer-R} to validate the effect of model design. \textbf{R-Transformer-V} is a vanilla R-Transformer model that does not apply the repetition learning matrix or rule-based generation. \textbf{R-Transformer-R} is a vanilla R-Transformer model that does not apply rule-based generation but applies the repetition learning matrix. In addition, \textbf{R-Transformer-RR} is our R-Transformer model that applies both the repetition learning matrix and rule-based generation. We let each model generate motif-level results using the given motif condition retrieved from 100 test samples. Specifically, 1-bar polyphonic piano music is generated based on the original motif. We evaluate the result using the matching rate $\mathcal{M}$ between output and input based on the definitions of different repetition types,
$$\mathcal{M} = \dfrac{\# \ \text{of matched motifs}}{\# \text{of total motifs}}$$
When two motifs satisfy the definition in Table \ref{tab:repetition}, we consider that a match; otherwise, the motifs do not match.

It should be noted that the goal of the proposed model is different from other models since this work tries to generate music by modeling music repetitions while other models do not necessarily generate different music repetitions since they are not designed to learn repetitions. Thus, to have a fair comparison, in the subjective evaluation, we report the overall music quality with other models in the paper since the ultimate goal of our model and other music generation models is generating enjoyable music.

\subsection{Analysis of Generated Examples}

\begin{figure}[!t]
	\centering
	{\centering\includegraphics[width=0.9\linewidth,trim=20 100 30 140,clip]{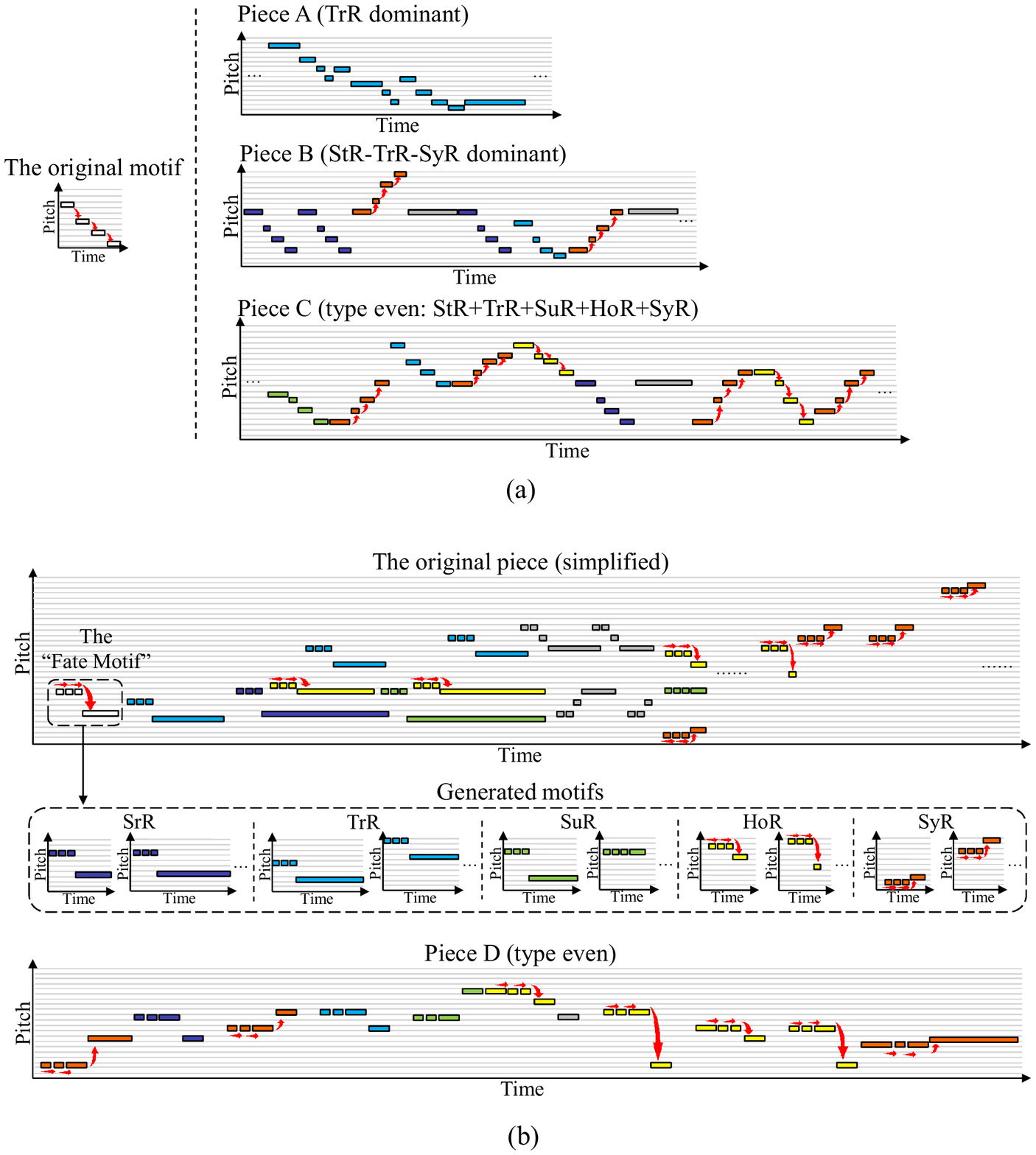}}
	\caption[.]{Examples of generated results. (a) The example is generated from the test dataset. (b) The example is generated from the ``fate motif''. Purple: strict repetition (StR). Blue: transpositional repetition (TrR). Yellow: subsequential repetition (SuR). Green: homodirectional repetition (HoR). Orange: symmetric repetition (SyR). Gray: ornamentation note \footnotemark[3]. Melodies are illustrated.}	
	\label{example}
	\vspace{-10pt}
\end{figure}

\begin{table*}[htb]
	\setlength\tabcolsep{12pt}
	\centering
	\caption{Objective evaluation results in terms of the matching rate.}
	
	\vspace{-10pt}
    \begin{tabular}{lccccccc}
	
		\toprule

		Models&StR&TrR&SuR&HoR&SyR\\ \hline
		\midrule

CP-C \cite{hsiao2021compound}&
0.00$\pm$ 0.00&
0.04$\pm$ 0.08&
0.21$\pm$ 0.41&
0.68$\pm$0.47&
0.71 $\pm$0.46&
\\
CP-NC \cite{hsiao2021compound}&
0.01$\pm$ 0.07&
0.11$\pm$ 0.31&
0.36$\pm$ 0.48&
0.62$\pm$0.49&
0.64 $\pm$0.48&
\\

PopMNet \cite{wu2020popmnet} &
0.00$\pm$ 0.00&
0.00$\pm$ 0.00&
0.10$\pm$ 0.30&
0.60$\pm$0.49&
0.70 $\pm$0.45&
\\

Theme \cite{shih2022theme} &
0.00$\pm$ 0.00&
0.00$\pm$ 0.00&
0.48$\pm$ 0.18&
0.47$\pm$0.25&
0.47 $\pm$0.25&
\\

MELONS \cite{zou2021melons} &
0.20$\pm$ 0.40&
0.20$\pm$ 0.40&
0.40$\pm$ 0.49&
0.70$\pm$0.46&
0.70 $\pm$0.46&
\\

R-Transformer-V&
0.54$\pm$ 0.49&
0.79 $\pm$0.41&
0.74$\pm$ 0.44&
0.85$\pm$0.36&
0.74 $\pm$0.42&
\\
R-Transformer-R&
0.84 $\pm$0.36&
0.96 $\pm$0.18&
\textbf{0.75}$\pm$ \textbf{0.43}&
0.95 $\pm$0.17&
0.75 $\pm$0.44&
\\
R-Transformer-RR &
\textbf{1.00 }$\pm$\textbf{0.00}&
\textbf{1.00} $\pm$\textbf{0.00}&
\textbf{0.75}$\pm$ \textbf{0.43}&
\textbf{0.97}$\pm$ \textbf{0.17}&
\textbf{0.75}$\pm$ \textbf{0.43}&
\\
		\bottomrule
	\end{tabular}
	\label{quant}
\end{table*}

\footnotetext[3]{Ornamentation note: the addition of notes for expressive and aesthetic purposes.}
To validate the proposed model can not only generate diverse and beautiful repetitions, but also achieve pleasant and complex melodies, we demonstrate two examples that is displayed in Figure \ref{example}. In the first example, the original motif is made up of four consecutive notes (G\#-F-D\#-C\#) with a downward melodic movement. Piece A is a TrR dominant version, which delivers a feeling of ``pacing back and forth'' or ``jumping over and over''. Although the intervals within each motif are the same, TrR expands the pitch distribution of the melody, causing the oscillation of same pattern at different pitch levels. Piece B is a StR-TrR-SyR dominant version, which creates melodic waves and emotional ups and downs. The introduction of StR helps to strengthen the original pattern, while SyR brings a distinct melodic direction and richer intervals. Therefore, this piece is no longer an one-way repetition with the descending direction, but intersperses with a number of upward trends. Piece C is a type even version that applies all five repetition types, which is more coherent and exquisite in melodic and emotional changes. SuR offers slight changes to the original motif, bringing familiarity and freshness. HoR contains a variety of motifs with the same direction of movement and shifting intervals, making the music unrestrained and diverse. Through exquisite combinations, the music expresses complex consciousness and emotions.

The second example is generated based on the ``fate motif''. By comparing similarities and differences between the generated motifs and motifs in Beethoven’s Fifth Symphony, we can further verify that the rich motif variants can form diverse and beautiful musical pieces with high quality. First, the model can generate all kinds of variants of the ``fate motif'' that appear in the original piece. Motifs in the middle row of Figure \ref{example} (b) mainly appear in the early stage of the first movement, which is the most recognizable part of Beethoven's Fifth Symphony. The appearance of these motifs lays the emotional tone and presents the dominant theme for the entire first movement and even the whole symphony. Therefore, by selecting and combing the generated motifs, we can reproduce a piece of music that is similar to the original work, and even make the generated music express similar feelings. Second, Piece D is a type-even version generated from the ``fate motif'' that delivers distinct feelings compared to the original Beethoven’s Fifth Symphony. Overall, Piece D presents a relaxed, leisurely and humorous feeling. Specifically, some adjacent motifs with opposite directions of movement naturally form some pairs, creating a feeling of ``walking together'' or ``asking and answering''. The development of the whole music is gentle without huge jumps and turns. All the differences are generated from one motif, but transfer the Beethoven’s Fifth Symphony from tragic and passionate to relaxed and joyful, which further confirm the richness and beauty of repetition. Listening samples can be found in supplementary materials.

\subsection{Objective Evaluation}

Table \ref{quant} lists the matching rate of five music generation models and three variants of the proposed model when generating one bar. Overall, the proposed model produces a good performance when generating different repetition types. 

Upon comparing the results of the proposed model with other music generation models in terms of repetition generation, the proposed model outperforms all other models in all repetition types since the proposed model is designed to recognize different repetitions. In addition, HoR and SyR are higher than StR, TrR and SuR in other models. The reason is that StR, TrR and SuR have a more strict definition and these models tend to learn those that appear the most or the easiest to learn since there is no supervisory signal for repetitions. This result also confirms the deficiency of current models.

It is also noted that R-Transformer-R is more robust than R-Transformer-V after applying the repetition learning matrix. It is reasonable since repetition learning matrix tells the model which attributes are more important to a specific repetition type during training. A comparison of R-Transformer-RR with R-Transformer-R reveals that StR and TrR achieve a matching rate of 1. StR and TrR follow explicit rules based on the definition; intuitively, implementing rules to generate pitch will lead to a matching rate of 1.

\subsection{Subjective Evaluation}
To validate that different repetitions can actually create pleasant music, we also design a subjective evaluation on Chinese social media. 150 subjects were randomly recruited on the internet. 27 reports were detected invalid because of some reasons such as submission too fast, no response, random answer, etc. The remaining 123 reports were used for analysis. None of the subjects has any prior knowledge of our study. Among the 123 participants,  57 were men and 66 were women; 65 were under age 25, 46 were between 25 and 50, and 12 were older than 50. In addition, 44 had musical training and were familiar with music theory (labeled ``pro''); the others possessed no musical training and were labeled ``non-pro''. 

In the experiment, we let subjects listen to seven sets of music: six sets of music are generated by models and one by human composers, which is the original music from the test dataset based on the selected motif. In all sets, subjects are asked to listen to two pieces of music, namely an original motif and a piece comprising several motifs. In R-Transformer set, the music piece is combined with different types of repetitions generated by the model. Subjects randomly select the music in each set; none indicated having heard the original motif before the test. During the experiment, each subject is asked to rate the overall quality on a 5-point scale (from 1 to 5; the higher the better) after each piece of music.

\begin{figure}[h]
    \vspace{-5pt}
	\centering
	{\centering\includegraphics[width=0.95\linewidth,trim=0 100 90 90,clip]{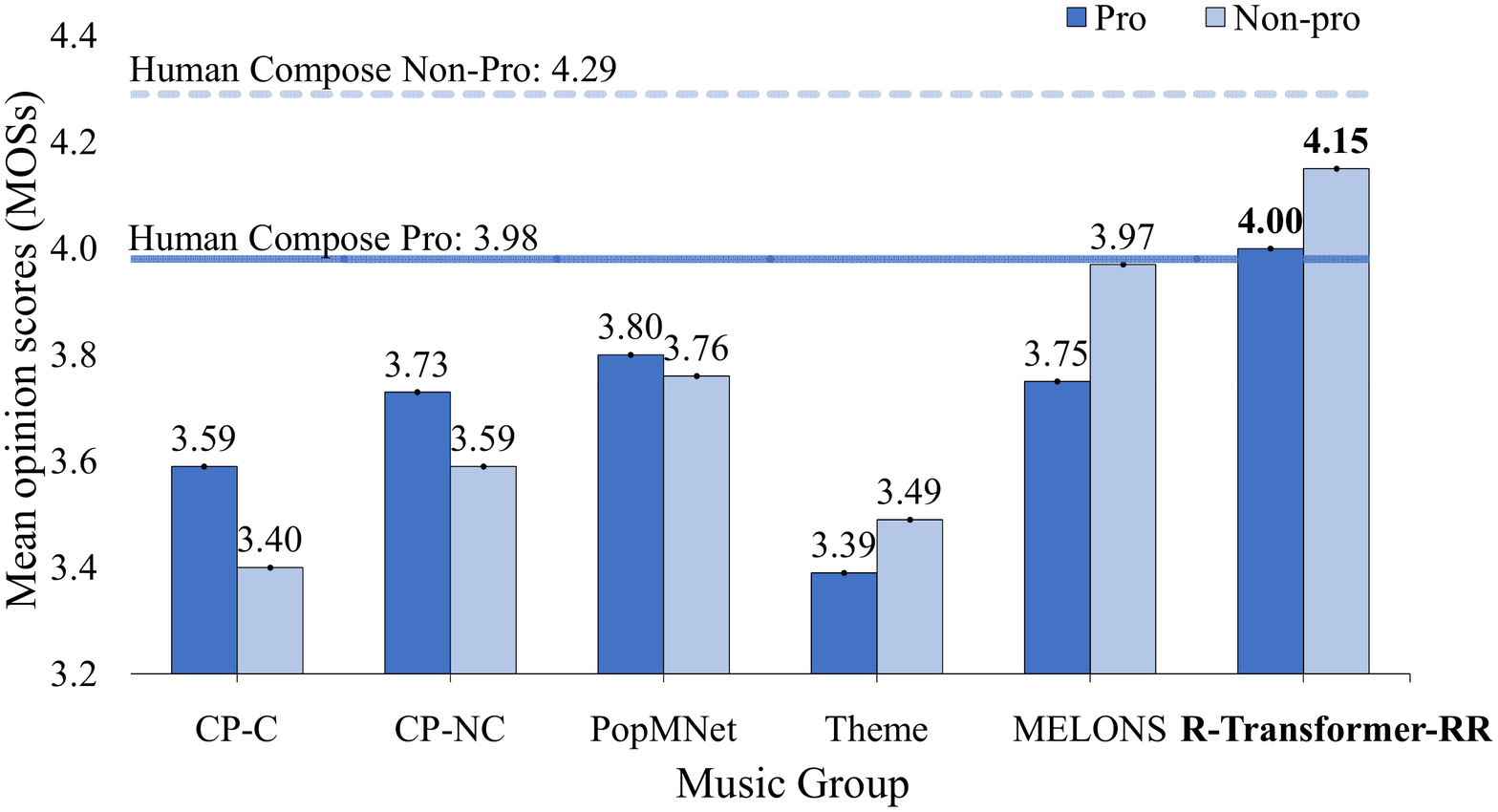}}
	\vspace{-10pt}
	\caption[.]{Subjective evaluation in terms of quality.}	
	\label{quality}
	\vspace{-5pt}
\end{figure}

Figure \ref{quality} shows the mean opinion scores (MOSs) on the overall quality in the ``Pro'' and the ``Non-pro'' group. The results demonstrate the effectiveness of the proposed method. In terms of overall quality, the proposed model outperforms other models in both groups, indicating that the proposed model can generate pleasant music based on different repetition types. Surprisingly, the proposed model performs better than human-composed music in the ``Pro'' group, verifying our model's authenticity. These strengths likely emerged for two reasons. First, by combining different repetition types, the results of R-Transformer sound vivid and structural because the entire music piece is developed from one motif. Second, repetition is confirmed to be an important factor that makes the music pleasant.

Table \ref{ttest} shows a paired t-test of all 123 subjects between the human-composed music and the machine-composed music, which is used to evaluate whether differences exist between two variables for the same subject. There is no significant difference between the music generated by the proposed model and the original human-composed music (p=0.351), while there is a significant difference when comparing results generated by other models and original human-composed music (p<0.05), indicating a better music quality of the proposed model. 

\begin{table}[htb]
    \vspace{-5pt}
	\setlength\tabcolsep{10pt}
	\centering
	\caption{Paired t-test results between machine-composed music and human-composed music.}
	\begin{tabular}{lccc}
		\toprule

		Music Group&Mean&Std&T-test\\ \hline
		\midrule
CP-C \cite{hsiao2021compound}&
3.47&
1.15&
t=-6.01, p<0.001\\
CP-NC \cite{hsiao2021compound}&
3.64&
0.96&
t=-4.74, p<0.001
\\
PopMNet \cite{wu2020popmnet} &
3.77&
0.95&
t=-4.41, p<0.001
\\
Theme \cite{shih2022theme} &
3.46&
0.91&
t=-7.37, p<0.001
\\
MELONS \cite{zou2021melons} &
3.89&
1.85&
t=-3.43, p=0.001
\\
The proposed&
4.10&
0.95&
t=-0.94, p=0.351
\\
Human-composed&
4.18&
0.84&
-
\\
		\bottomrule
	\end{tabular}
	\label{ttest}
    \vspace{-5pt}
\end{table}

\section{Conclusion}
To the best of our knowledge, this is the first comprehensive study of repetition in machine composition. We construct a dataset named MRD based on the Pop piano dataset \cite{hsiao2021compound}. 562,563 training data and 21,766 test data are provided with the labels of motif level repetitions. A novel repetition generator is designed based on the transformer encoder and a repetition-aware learner. The proposed R-Transformer can generate a large amount of motif-level repetition of different types. Moreover, the learned composition skills on Pop music dataset also demonstrate very interesting results on fate motif from the classical music. Subjective evaluation on both  musicians and non-professional users show that the proposed techniques help to improve the quality of the machine composed music obviously. 

\section{Acknowledgments}
The authors would like to thank the reviewers for their constructive comments. The work is supported by The Hong Kong Polytechnic University for the project: Artificial Intelligence and Robotics (AIR) Group: Artificial Intelligence Art under Grant No.: P0033738.

\newpage
\balance
\bibliographystyle{ACM-Reference-Format}
\bibliography{samplebase}

\end{document}